\documentclass[10pt,twocolumn]{article}
\usepackage{graphicx}
\usepackage{graphicx,amsmath,amssymb}

\newcommand{\Section}[1]{\vspace{-8pt}\section{\hskip -1em.~~#1}\vspace{-3pt}}

\begin{document}

\twocolumn[
{\LARGE \bf Modeling the Envelope of the Secular Light Curve of the Comet 1P/Halley } 

 \vskip10mm

 \begin{raggedright}
{\small \centering{{\bf Eduardo Rond\'on} (1,2) and Ignacio Ferr\'in (2)}\\
\vskip5mm
(1) N\'ucleo Alberto Adriani, Facultad de Ingenier\'ia, Universidad de Los Andes, M\'erida, Venezuela, (2) Centro de F\'isica Fundamental, Facultad de Ciencias, M\'erida, Venezuela (erondon@ula.ve ).} 
\end{raggedright}

\vskip10mm %
]

\thispagestyle{empty}

\section*{Abstract} 

We have been modeling the envelope of the secular light curve (Ferrín, 2010)\cite{Ferrin10} of comet 1P/Halley. As the first step we have modeled the water production rate of the comet, for several active regions as a function of visual albedo and infrared albedo, and then we have used the correlation between the water production rate and the reduced magnitude (Jorda, 2008)\cite{Jorda08} for modeling the envelope of the secular light curve of the comet. We obtain probable orientations of the rotation axis (I,$\Phi$). These orientations are compared with several solutions by several authors. We have calculated the surface temperature for the orientation of the rotation axis with minimal standar deviation. We have found that for  near and far values of the perihelion the surface temperature as a function of the latitude is constant. 

\Section{Introduction} 

Comet Halley has a large number of observed apparitions, and many visual observations have been published in the literature. Ferr\'in (2010)\cite{Ferrin10} has published brightness observation of several comets, considering that the envelope of the data describes the behavior of the brightness in time. Several secular light curves show asimmetry. We have developed a model that  predicts this asimmetry. The orientation of the rotation axis explains the asimmetry in the secular light curve (Rond\'on \& Ferrín, 2010; Rond\'on, 2007)\cite{Rondon10}\cite{Rondon07}. The importance of studying the secular light curve is that it gives a large amount of physical informations of comet. By modeling this curve we can explain the behavior of the brightness mathematically and we can predict the physical parameters of the curve. 



\Section{Model Calculations}

The first step for modeling the light curve is to calculate the sublimation rate of water. The equation that describes the vaporization rate of a comet is the energy conservation equation, given by:

\begin{eqnarray}
F_{0}(1-A_{v})r_{H}^{-2}\overline{cos(\theta)}= (1-A_{ir})\sigma T^{4}+ \nonumber \\ 
\hspace{2cm} Z(T)L(T)+ K \frac{\partial(T)}{\partial(z)}
\end{eqnarray}

Where A$_{ir}$ is the infrared albedo, r$_H$ Sun-comet distance, $\overline{cos(\theta)}$ is the projection factor (Cowan\&A'Hearn, 1979)\cite{Cowan79}, $\sigma$ is the Steffan Bolztmann constant, T  is the temperature, Z(T) is the sublimation function, L(T) is the latent heat function, K is the thermal conductivity constant, z is the layer depth.

\begin{equation}
Z(T)=\frac{P(T)m}{2(\pi)kT}
\end{equation}
where $P(T)$ is the  vapor pressure function, m is the molecular weight, k is the ideal gas constant.

If  we know the vapor pressure function and the latent heat function, we can solve for the  energy  conservation equation(1).
\begin{equation}
Z_{total}=\overline{Z(i)}=\frac{1}{2}{\int_{-\pi/2}^{\pi/2} (Z(i,b)cos(b)db)}
\end{equation}

We have modeled the water production rate for comet 1P/Halley using the observational data given by (Schleicher, 1998)\cite{Schleicher98}, and we have found that when considering an active region of 80 km² we can predict the average of the observational data (Fig.1a). However, the secular light curve considers the envelope of the data (Ferrin, 2005)\cite{Ferrin05}. For this case we have taken an active region of 180 km²(Fig.1b). In both cases we have assumed an visual albedo $A_v=0.009$, an infrared albedo $A_ir=0.5$ and a thermal conductivity K=0.

The correlation equation between the reduce visual magnitude and the water production rate is known (Jorda, 2008)\cite{Jorda08} and is given by:

\begin{equation}
 m=125.051-4.077\log(Z_{total})
\end{equation}
\\
The secular light curve of comet Halley is assymetric. This can be explained through of the orientation of the rotation axis of comet (Rond\'on, 2007)(Rond\'on \& Ferr\'in, 2010)\cite{Rondon07}\cite{Rondon10}. 

\begin{figure}[!ht]
\vspace{-0.1 cm}
\begin{center}
\includegraphics[width=2.7 in]{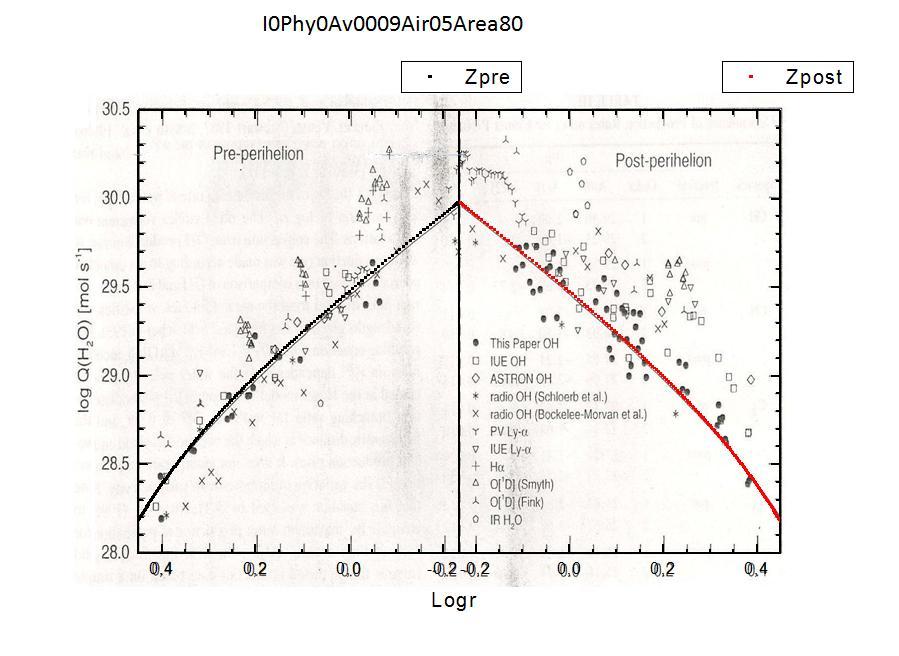}
\vskip-5mm
\includegraphics[width=2.7 in]{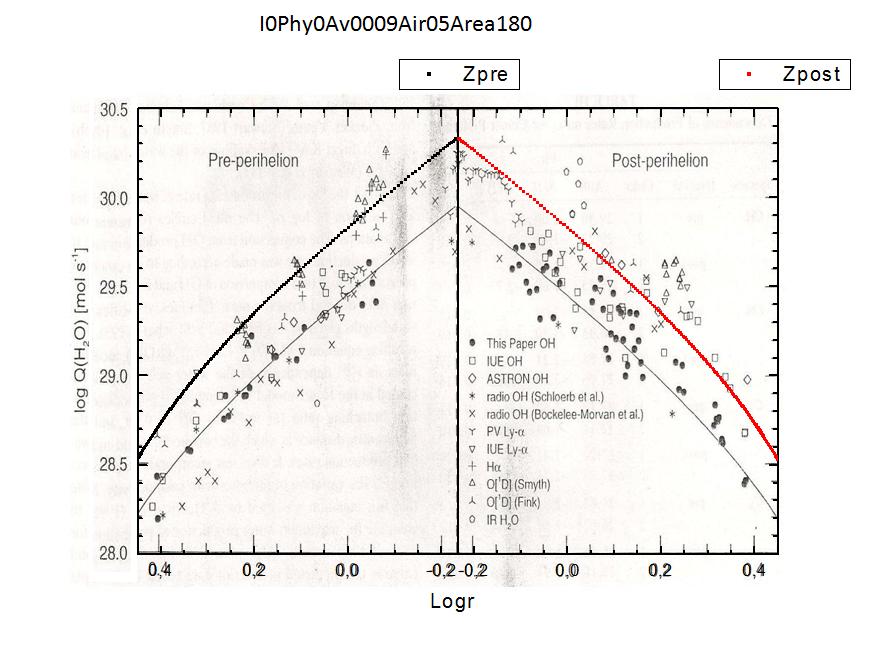}
\vskip-5mm
\caption{Water production rate vs log(r), a) with an active region of 80 km² b) with an active region of 180 km².}
\end{center}
\end{figure}

\begin{figure}[!ht]
\vspace{0.2 cm}
\begin{center}
\includegraphics[width=2.5 in]{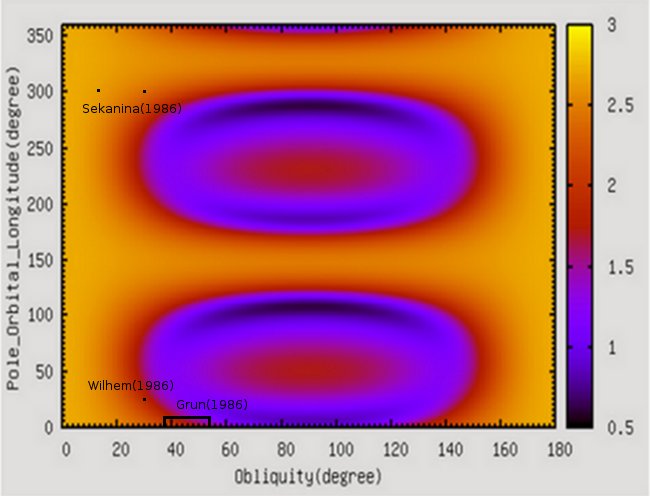}
\caption{Standard deviation for each of the orientation of the rotation axis for the comet 1P/Halley.}
\label{fig2}
\end{center}
\end{figure}
\vskip -1.5mm

We have calculated the standar deviation for our model with the observational data, and found that the solution with minimal standar deviation is for a obliquity, I = 90° and a pole orbital longitude, $\Phi$ = 112°, (Fig, \ref{fig2}). Several authors obtained solutions for the orientation of the rotation axis (Sekanina, 1986) \cite{Sekanina86}.

\begin{figure}[!ht]
\vspace{-0.3 cm}
\begin{center}
\includegraphics[width=2.7 in]{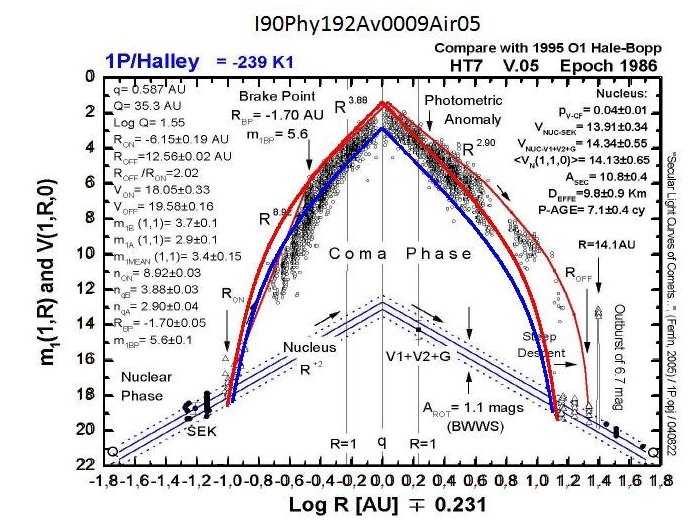}
\vskip -3mm
\caption{Secular light curve of the Halley comet with a orientation of the rotation axis of I = 90°, $\phi$ = 112°. In the upper envelope curve has been considered an active region of 180 km² and in the lower envelope curve has been considered an active region of 80 km².}
\label{fig3}
\end{center}
\end{figure}
\vskip-7mm
\
\\
In Figure \ref{fig2} we can see that a region of the solution of (Grun, 1986)\cite{Grun86} has a standard deviation of only 1 magnitude. In the Figure \ref{fig3} we have modeling the secular light curve assuming an orientation of the rotation axis of I= 90°, $\Phi$=112° for an active region of 80 km²(blue line), and 180 km², (red line).	In this graph we can see that before perihelion the model fits the observational data, but after perihelion the fits  is not very good. This behavior is due to the thermal conductivity effect in the nuclear surface that has not been considered in the model.
\
In Figure 4  we can see that the surface temperature as a function of the latitude is constant for values of the heliocentric distance near and very far away of perihelion. This behavior near of perihelion is that the coma of comet tends to stabilize the temperature, while for heliocentric distance very far away perihelion the sun radiation is small and therefore the whole cometary nucleus cools.    
\
\\
\\
\begin{figure}[!ht]
\vspace{-0.3 cm}
\begin{center}
\includegraphics[width=2.5 in]{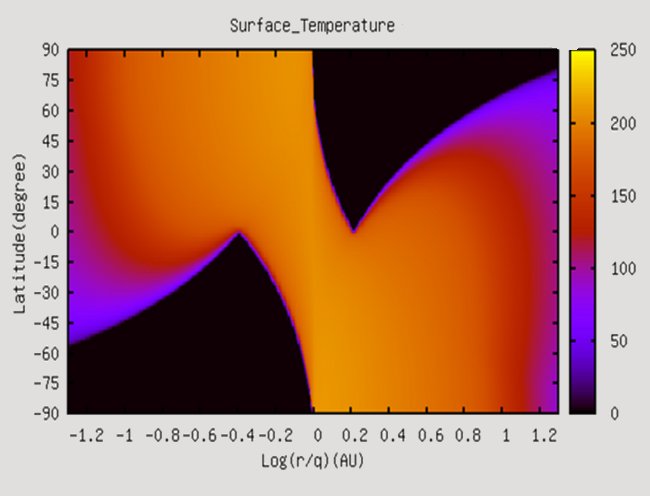}
\vskip 1mm
\caption{Surface temperature as a function of heliocentric distance, assuming an active region of 180 km² and a orientation of the rotation axis of I= 90°, $\Phi$ = 112°.}
\label{fig4}
\end{center}
\end{figure}
\
\Section{Summary and Conclusions}

We have developed a theoretical model capable of reproducing the observational data of the secular light curve and water production rate of comet Halley, with minimal standar deviation $\sigma$ = 0.3 mag,  for an orientation of the rotation axis I = 90°, $\phi$ = 112°. The region of the solution  shown by (Grun, 1986)\cite{Grun86} explain the secular light curve with a standard deviation $\sigma$=0.6	5 mag. We also have found that the surface temperature as a function of the latitude is constant for values of the heliocentric distance near and very far away of perihelion. \

\end{document}